\documentclass[usegraphicx]{mn2e}
\usepackage{times}
\usepackage{graphicx,subfigure}

\begin{document}
\newcommand{\laboca}{LABoCa}

\title[Submillimetre Variability of Eta Carinae]{Submillimetre Variability of Eta Carinae: cool dust within the outer ejecta}

\author[H.L. Gomez et al.]{H.L. Gomez$^{1}$\thanks{E-mail: haley.gomez@astro.cf.ac.uk},  C. Vlahakis$^{2}$, C.M. Stretch$^1$, L. Dunne$^{3}$, S.A. Eales$^1$, A. Beelen$^{4}$, \and E.L. Gomez$^{5,1}$, M.G. Edmunds$^{1}$\\
  $^{1}$ School of Physics \& Astronomy, Cardiff University, The Parade, Cardiff CF24 3AA, UK\\
  $^{2}$ Leiden Observatory, Leiden University, PO Box 9513, 2300 RA Leiden, The Netherlands \\
  $^{3}$ School of Physics \& Astronomy, University of Nottingham, University Park, Nottingham, NG7 2RD, UK\\
$^{4}$ Institut d’Astrophysique Spatiale, b\^{a}t 121, Universit\'{e} Paris-Sud, 91405, Orsay, Cedex, France \\
  $^{5}$ Las Cumbres Observatory Global Telescope Network, Inc., 6740 Cortona Dr, Suite 102, Goleta, CA 93117, USA}



\maketitle

\begin{abstract} 
  Previous submillimetre (submm) observations detected $\rm
  0.7\,M_{\odot}$ of cool dust emission around the Luminous Blue
  Variable (LBV) star $\eta$ Carinae. These observations were hindered
  by the low declination of $\eta$ Carinae and contamination from
  free-free emission orginating from the stellar wind.  Here, we
  present deep submm observations with \laboca~at 870\,$\mu$m, taken
  shortly after a maximum in the 5.5-yr radio cycle.  We find a
  significant difference in the submm flux measured here compared with
  the previous measurement: the first indication of variability at
  submm wavelengths.  A comparison of the submm structures with
  ionised emission features suggests the 870\,$\mu$m is dominated by
  emission from the ionised wind and not thermal emission from dust.
  We estimate $0.4\pm 0.1\,\rm M_{\odot}$ of dust surrounding $\eta$
  Carinae.  The spatial distribution of the submm emission limits the
  mass loss to within the last thousand years, and is associated with
  mass ejected during the great eruptions and the pre-outburst LBV
  wind phase; we estimate that $\eta$ Carinae has ejected $>\rm
  40\,M_{\odot}$ of gas within this timescale.
\end{abstract}


\section{Introduction}

$\eta$ Carinae ($\eta$ Car), one of the most luminous infra-red (IR)
objects in our Galaxy ($\rm > $$10^6L_{\odot}$), is well known for
dramatic outbursts over the last thousands of years, in which vast
amounts of material is ejected outwards from the star.  The great
eruption phase during the last 200 years are responsible for creating
the dusty bipolar nebula known as the Homunculus, as seen in the {\it
  Hubble Space Telescope (HST)} images.  The recent discovery of a
fast blast wave from the 1843 eruption (Smith 2008) indicates that
the mass loss in $\eta$ Car is not simply due to a stellar wind, but
could be more akin to a low-energy supernova remnant. Extensive IR
imaging has revealed important clues about the mass loss history; the
Infrared Space Observatory (ISO) discovered a massive cool dust
component (Morris et al.\ 1999; Smith et al.\ 2003) with a total of
$\sim$$\rm 0.1\,M_{\odot}$ of dust proposed to be present in the
nebula, suggesting a mass loss rate of $0.5\,\rm M_{\odot}$\,yr$^{-1}$
during the last 200 years.  Submillimetre (submm) observations (Gomez
et al.\ 2006, hereafter G06) of $\eta$ Car at 450 and 850\,$\mu$m with
the Submillimetre Common User Bolometer Array (SCUBA) detected the
  presence of a massive component of dust, with $\sim$$\rm
  0.7\,M_{\odot}$ needed to reproduce the IR-submm Spectral Energy
  Distribution (SED).  Their work suggested that up to four times more
  mass had been ejected from $\eta$ Car during its recent, violent
  history than previously proposed.  The SCUBA data also indicated
  that the cool dust component was extended along the mid-plane, far
  beyond the inner optical and IR region; this was interpreted as mass
  which had been ejected on a much longer timescale than the
  well-known, smaller features such as the torus and Homunculus.

  The previous analysis of the SCUBA data was hindered by two
  unknowns: whether the submm emission originated from free-free and
  not thermal emission from dust and whether the extended
  structure was reliable.  Firstly, the submm fluxes are likely to be
  contaminated by strong free-free emission from the extended stellar
  wind which is thought to vary with frequency as $\nu^{0.6-1.3}$
  (e.g. Cox et al. 1995; Brooks et al.\ 2005, hereafter B05). This is
  further compounded since the radio and millimetre (mm) fluxes are
  highly variable, with $\eta$ Car at 3\,cm changing appearance from a
  point source in 1992 to an extended region with five times more flux
  in 1996 (Duncan \& White 2003). The X-ray, radio and mm variability
  of $\eta$ Car has been well documented (Pittard et al.\ 1998;
  Abraham et al.\ 2005; Damineli et al.\ 2008), with the star
  undergoing periodic variability over a 5.5-yr period.  This
  variability has been associated with shocks from an extended disc
  surrounding the $\eta$ Car binary system (e.g. Duncan \& White 2003)
  and/or due to episodic mass loss which, in turn, decreases the
  number of ionising photons. Since there were no available mm
  observations to match the epoch of the original 1998 SCUBA data
  (taken during the minimum phase of the radio cycle), G06 were not
  able to estimate the contribution to the submm due to free-free
  emission.  Second, the SCUBA data were taken at very large airmass
  ($A\sim 5.5$) due to the declination of $\eta$ Car.  This meant that
  the beam shape at 450\,$\mu$m was distorted due to the deformation
  of the dish at such low elevations and the morphology of the source
  was difficult to ascertain from this dataset.  Accurately
  determining the distribution and mass of dust in the stellar wind
  has enormous consequences not only on determining the mass loss
  history of $\eta$ Car, but also understanding the progenitors of
  `exotic' core-collapse supernovae (SNe; Pastorello et al.\ 2007;
  Smith et al.\ 2009).

  In this Letter, we present deep submm \laboca~observations of $\eta$ Car,
  taken shortly after the maximum phase in the cycle.  In
  \S\ref{sec:obs} we discuss the data reduction and compare the
  870\,$\mu$m emission to well-known features seen in X-ray, optical,
  H$\alpha$, 8\,$\mu$m and 1.2\,mm in \S\ref{sec:res}.  In
  \S\ref{sec:disc} we estimate the dust and gas mass for the
  stellar wind and our conclusions are presented in \S\ref{sec:conc}.

\section{Observations and Data Reduction}
\label{sec:obs}

The 870\,$\mu$m data were taken with the \laboca~camera (Siringo et
al.\ 2009), a 295-pixel bolometer array, located on the Atacama
Pathfinder EXperiment telescope (G\"{u}sten et al.\ 2006) on
Chanjantor in Chile. $\eta$ Car was observed during Science
Verification on 23rd July 2007.  This epoch corresponds to 1.4 yrs
after maximum brightness in the radio and X-ray cycles (2006.25).  The
observations were carried out in raster spiral mode with twelve scans,
each providing a fully sampled map over $30^{\prime} \times
30^{\prime}$.  The total on-source integration time was 1.7\,hours.
Two independant mesurements of the optical depth, $\tau$ were
obtained.  The first method used the precipitable water vapour (PWV)
levels measured every minute along the line of sight, then scaled
using the relevant atmopsheric transmission model. The PWV ranged from
0.7--0.9\,mm.  The second method calculated $\tau$ from skydip
measurements, where a model of the dependence of the effective sky
temperature on elevation were fitted to determine the zenith
opacity. The two skydips taken before and after the on-source scans
were both well fitted by the model, with $\tau$ between 0.1--0.2.
These values were 25\% lower than those estimated from the PWV
measurements.  A linear combination of the two methods, and a final
comparison with the calibrator models (e.g. Siringo et al.\ 2009), was
used to determine the values used in the data reduction.  Following
Dunne \& Eales (2001), the fractional error in the flux from the error
due to the range of opacities calculated with these two methods is
6\,per\,cent.

\begin{figure*}
\includegraphics[width=16cm]{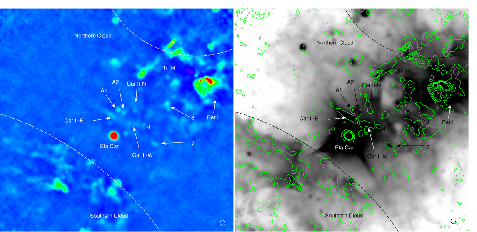}
\includegraphics[width=16cm]{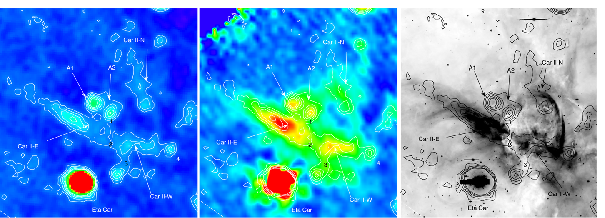}
\caption{\small{The environment surrounding $\eta$ Car.  {\bf Top
      Panel}: {\it Left:} \laboca~870\,$\mu$m emission.  {\it Right}:
    {\it MSX} 8\,$\mu$m emission, shown here in negative log
    greyscale, with \laboca~$870\,\mu$m contours (starting at
    0.04\,Jy\,beam$^{-1}$). Region shown covers $19 ^{\prime} \times
    19^{\prime}$ area with \laboca~beam size shown in lower right hand
    side.  Known compact sources are labeled 4-8 (B05).  {\bf Bottom
      Panel}: Zoomed in region ($9 ^{\prime} \times 8.4^{\prime}$)
    with \laboca~ 870\,$\mu$m contours overlaid, ranging from
    0.04--0.25\,Jy\,beam$^{-1}$.  {\it Left}: \laboca~emission. {\it
      Middle}: 1.2\,mm emission (SIMBA map kindly provided by K Brooks
    B05).  {\it Right:} H$\alpha$ emission shown here in negative
    greyscale (CTIO map kindly provided by N Smith, Smith, Bally \&
    Morse\ 2003).  The Keyhole Nebula is labeled 1, 2 and 3. }}
\label{fig:large}
\end{figure*}

The data were reduced using the BoA (BOlometer array Analysis
software) package.  The focus was checked on observations of Venus and
Jupiter and was stable within $\pm$0.2\,mm. Pointing observations were
within $1^{\prime \prime}$ in azimuth and elevation during the earlier
scans but crept up to within $3^{\prime \prime}$ towards the end.  Bad
and noisy pixels were flagged, with 276 bolometers used in each
scan. The data were despiked and correlated noise was removed.  The
reduction is optimized for the recovery of strong sources.  The scans
were coadded (weighted by rms$^{-2}$) and the data were gridded onto
$6^{\prime \prime}$ pixels to create the final map. After a first
iteration of the reduction, the source map was used to flag bright
sources and the data were reduced again. This was efficient at
removing negative artifacts which appear around the bright sources in
the first iteration and led to a more stable background noise level in
the central region ($\sim$40\,mJy\,beam$^{-1}$).  The 870\,$\mu$m map
towards $\eta$ Car is shown in Fig.~\ref{fig:large} together with
8\,$\mu$m {\it Midcourse Space eXperiment} ({\it MSX}) archival data.
$\eta$ Car is the strongest source in the \laboca~map, and peaks at
signal-to-noise $>$100.  Well known emission regions, molecular clouds
and star clusters are identified in the image.

Secondary calibrators were used to calibrate the map and the
typical correction factor when comparing the measured flux on the
calibrator with its expected flux in an aperture was
$\sim$5\,per\,cent. The total uncertainty in the calibration,
including uncertainties in the calibrator model is therefore
$\sim$12\,per\,cent. The 870\,$\mu$m flux of $\eta$ Car measured in an
aperture with radius $30^{\prime \prime}$ is $42\pm
5$\,Jy, compared to the flux estimated by SCUBA in 1998 of $13\pm
1$\,Jy (G06). This is an increase by a factor of three over the 9\,yr period
and cannot be accounted for by calibration errors.

\section{Results}
\label{sec:res}

A zoomed in image of $\eta$ Car is presented in Fig.~\ref{fig:large}
and compared with H$\alpha$ (Smith, Bally \& Morse.\ 2003) and 1.2\,mm
(B05) emission.  Faint structures not previously seen in FIR-submm
maps are detected with \laboca~and are identified with well
known sources (labeled in Fig.~\ref{fig:large}
following the notation of Smith et al.\ 2003 \& B05). The submm
emission can be separated into four components: the bright source at
the location of $\eta$ Car; filaments and arcs which surround
molecular globules (Rathborne et al.\ 2002); peaks and filaments
associated with molecular clouds at the edges of the map; and a number
of compact sources (seen at mm and submm wavelengths, but not always
in H$\alpha$ and 8\,$\mu$m). The edges of the two large molecular
clouds, part of a giant cloud complex within the Carina arm, are
visible in the \laboca~map to the North and South of $\eta$ Car. The
870\,$\mu$m flux in these regions correlates with peaks seen in
molecular emission (Cox et al.\ 1995; Yonekura et al.\ 2005)
suggesting dust emission from cold clouds.

$\eta$ Car is the brightest source at 870\,$\mu$m and 1.2\,mm, but not
at IR-FIR.  At 12 and 25\,$\mu$m, the peak is found at the
centre of Tr 14 (a massive star cluster) whereas at 60 and
100\,$\mu$m, the peak shifts towards the west (and the Car I
feature). This shift traces a temperature gradient in the nebula (Cox
1995); hotter dust is located closer to Tr 14, where massive stars are
externally heating the clouds and cooler material is forced further
away.  The shift in peak emission from IR-FIR is also seen in our
submm data.

The faint structure just north of $\eta$ Car is associated with the
Car II emission feature. The 870\,$\mu$m emission closely follows the
H$\alpha$, 8\,$\mu$m and 1.2\,mm emission (Fig.~\ref{fig:small}) and
at longer wavelengths, the 3\,cm radio emission (B05).  These arcs and
filaments surround dark clouds detected in CO emission (Rathborne et
al.\ 2002, Cox \& Bronfman 1995).  The close correlation between
ionised gas, emission at both mm and radio wavelengths and the
\laboca~emission suggest that the 870\,$\mu$m originates from ionised
material.  Indeed, the integrated flux of Car II at 870\,$\mu$m here
agrees with the extrapolated radio continuum spectrum in B05 (their
Fig 5) which are well fitted with an H{\sc ii} emission model.  The
ratio of the 870\,$\mu$m/1.2\,mm maps confirm that the spectral index
in Car II varies as $F_{\nu} \propto \nu^{-2}$, consistent with an
ionization front or H{\sc ii} region, thus the \laboca~emission here
orginates from a free-free source and not from warm dust.  Our lack of
submm detection at the location of the famous Keyhole Nebula (labeled
1-3, Fig.~\ref{fig:large}) rules out any dust component with
temperatures higher than 13\,K.


\subsection{Submillimetre Emission from $\eta$ Carinae}
\label{sec:subeta}

\begin{figure}
\includegraphics[width=8.3cm]{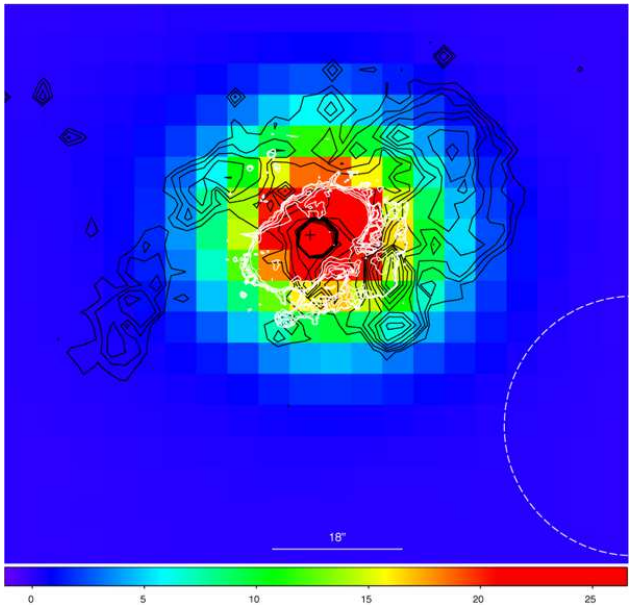}
\caption{\small{\laboca~ 870\,$\mu$m emission of $\eta$ Car with
    archival X-ray ({\small CHANDRA}) and optical ({\it HST}) contours
    shown in black and white respectively. The \laboca~beam
    ($19^{\prime \prime}$) is shown in the lower right hand side.
    Units are Jy\,beam$^{-1}$. The extent of the
    Homunculus seen as the double-lobe structure in the contours is
    $\sim 18^{\prime \prime}$. }}
\label{fig:small}
\end{figure}

The central $\eta$ car source as seen with
\laboca~(Figs.~\ref{fig:large} \& \ref{fig:small}) is well fitted by a
two-dimensional Gaussian with {\sc fwhm} $22^{\prime \prime}$.  We
find that $\eta$ car is unresolved with the \laboca~beam ($19^{\prime
  \prime}$) and we see the first error beam pattern below 1\,per\,cent
peak intensity at 0.3\,Jy\,beam$^{-1}$. Deconvolving from the beam and
pixel size indicates a central source with {\sc fwhm}
$\sim$$10^{\prime \prime}$.  We do not detect the structure along the
midplane seen in the original 850\,$\mu$m data from G06, confirming
the suggestion made there that a combination of high airmass and chop
throw had smeared the signal across the midplane.  Based on the
\laboca~observations, the true extent of the submm source is likely to
be within the Homunculus ($18^{\prime \prime}$) and interior to the
region encompassed by the O\,{\sc III} `cocoon' of material
($60^{\prime \prime}$) detected by Smith, Morse \& Bally (2005).
Recently, Smith (2008) detected a fast blast wave, traveling at up to
6000\,km\,s$^{-1}$, from the great eruption in 1843. This blast wave
extends out to the same radius as the cocoon, and is interior to or
coincident with, the X-ray emission.  Smith proposed that the X-rays
arise from this fast blast wave running into a previous eruption which
gave rise to a nitrogen rich shell.  Such a scenario places severe
constraints on the timescales for emission seen interior to the blast
wave material and suggests that the source responsible for the
emission seen by \laboca~was ejected within the last 200--1000\,years.  

The 1.2\,mm emission from $\eta$ Car is closely correlated with the
870\,$\mu$m, and neither correlate well with the IR-FIR central
source.  If we assume that the former originates from thermal emission
from dust grains (i.e. $F_{\nu} \propto \nu^{3+\beta}$), the ratio of
870\,$\mu$m/1.2\,mm would require a dust emissivity index, $\beta$, of
-2.  If we assume emission at both wavelengths originates from a
source which varies as $F_{\nu} \propto \nu^{-\alpha}$, we obtain
$\alpha=0.5$--$2$ which agrees well with emission from an ionised
stellar wind source.  The similar variability seen at both 870\,$\mu$m
and 1.2\,mm suggests that less than 20\,per\,cent of the 870\,$\mu$m
flux is contributed by dust emission.

\section{Discussion}
\label{sec:disc}

\subsection{The Spectral Energy Distribution of $\eta$ Car}
\label{sed:discdust}

\begin{figure}
\includegraphics[angle=-90,width=8.5cm]{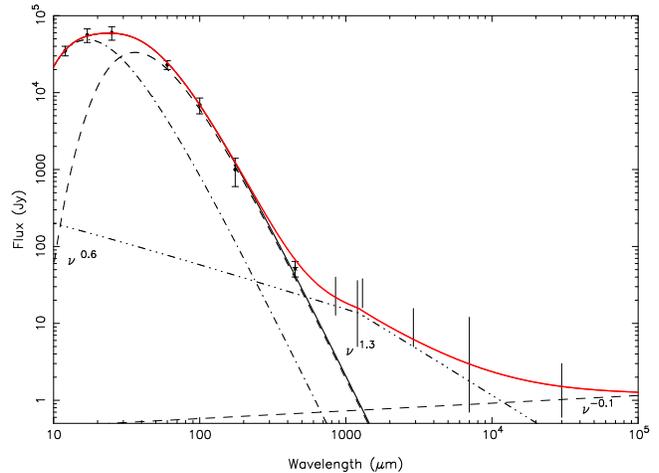}\hfill
\caption{\small{The SED of $\eta$ Car from 12\,$\mu$m through to the
    radio. The radio, mm and 850/870\,$\mu$m fluxes are shown as
    vertical bars (following B05) to indicate their variability range
    during the 5.5-yr cycle.  The solid black line is the sum of the
    two-temperature components fitted to the IR ($T_h = 174$\,K and
    $T_w = 82$\,K) and 450\,$\mu$m fluxes (dot-dashed and dashed
    respectively).  The dot-dot-dot-dashed line shows the SED expected
    from the optically thick free-free wind, $F_{\nu} \propto
    \nu^{0.6}$ and the optically thin wind with $ \propto
    \nu^{1.3}$.  The dashed line represents emission from the
    optically thin Little Homunculus feature $\propto
    \nu^{-0.1}$. The red line shows the total SED from all the
    components. The far-IR luminosity from $\rm
    12$--$1000$\,$\mu$m is $\rm 1.6\times 10^{6}\,L_{\odot}$.}}
\label{fig:sed} 
\end{figure}

The spectral energy distribution (SED) of $\eta$ Car is shown in
Fig.~\ref{fig:sed}.  Given the variability of observed fluxes over
different epochs during $\eta$ Car's radio cycle, we represent the variability as vertical bars
following B05, highlighting the flux changes over the cycle rather
than the fluxes measured at different epochs.  

There are three major components to the SED: thermal emission from
dust grains (following a modified blackbody with power-law index
$\beta$), free-free emission from the ionised stellar wind and
free-free emission from the optically thin Little Homunculus feature,
seen close to the central star (Duncan \& White 2003). The stellar
wind material consists of two separate components: the optically thin
varying as $\nu^{1.3}$ (e.g. B05) and the optically thick varying as a
classical wind with $\nu^{0.6}$ (Wright \& Barlow 1975; Lamers \&
Cassinelli 1999). The optically thin component from the Homunculus is
produced by the ionisation of $\eta$ Car's stellar wind by the
UV-radiation field of the hot binary companion.  This produces
free-free emission which varies as $\nu^{-0.1}$ and relates to the
Little Homunculus and torus (Teodoro et al.\ 2008).


For the dust component, we used two-modified blackbodies:
\begin{equation}
S_{\nu}= N_h\nu^{\beta}B(\nu, T_h) + N_w\nu^{\beta}B(\nu,T_w) 
\end{equation}
  
where $N$ is the normalisation term, $\beta$ is the dust emissivity
index and $B$ is the Planck function at frequency $\nu$ and
temperature $T$.  The SCUBA 850 and \laboca~870\,$\mu$m fluxes are not
included in the fit to the thermal emission from dust.  The
450\,$\mu$m was corrected for free-free emission expected at this
wavelength using the predicted power-law variation $\nu^{0.6}$ (as
shown in Fig.~\ref{fig:sed}) where $\sim$35\,per cent of the
450\,$\mu$m 1998 flux could originate from free-free emission.  The
revised 450\,$\mu$m flux due to dust in the stellar wind is therefore
$\sim$30\,Jy.  This agrees with the results from the Balloon-borne
Large Aperture Submillimeter Telescope who observed $\eta$ Car
from 250--500\,$\mu$m during the maximum in 2006 (Hargrave
priv. comm., Hargrave et al.\ in prep).  Hargrave et al.\ find no
evidence of an increased flux at 500\,$\mu$m when compared to the
SCUBA 450\,$\mu$m observed during the minimum.  This supports the idea
that the 450\,$\mu$m flux measured with SCUBA is still dominated by
thermal emission from dust grains and does not undergo significant
variability. Note that the 450\,$\mu$m flux is key to the dust mass
estimation since this extra constraint forces the fit to have a lower
dust temperature and a steeper emissivity index than the previous published
models (Morris et al.\ 1999; Smith et al.\ 2003; B05).  We rule out a
cold dust component since we cannot obtain a modified blackbody fit to
the submm-mm fluxes. An adequate fit can be obtained using only the mm
fluxes but this would require a dust component with $T$$\sim$$4$\,K; it
is difficult to understand how such cold dust could survive the close
environment of $\eta$ Car.

The dust mass is estimated using Eq.~\ref{eq:dust_mass}:
\begin{equation}
  M_d={S_{\rm 450}~d^2 \over{\kappa_{450}}} \left( {N_h\over{B(\nu_{\rm 450}, T_h)}} + {N_w\over{B(\nu_{\rm 450}, T_w)}}\right)
\label{eq:dust_mass} 
\end{equation} 

where $\kappa_{\nu}$ is the dust-mass-absorption coefficient ($0.27\rm
\,kg^{-1}\,m^2$ for interstellar dust at 450\,$\mu$m, James et al.\
2002) and $d$ is the distance to the star.  The best-fit ($\chi^2$) to
the SED requires a total dust mass of $\rm 0.4 \pm 0.1 \rm
\,M_{\odot}$ and $\beta$$\sim$$1.9$ (indicative of normal interstellar
dust grains), $T_h$=$174^{+16}_{-8}$\,K and $T_w$=$80^{+10}_{-6}$\,K. The errors are estimated from the bootstrap
technique where 1000 fits to the fluxes were made; the 68\% confidence
intervals are quoted here. Previous authors required $\beta \sim 1$ to
fit the mid-IR SED but higher values are necessary to fit the
long-wavelength submm fluxes at 450 and 870\,$\mu$m.

The ``transition'' between the classical $\nu^{0.6}$ and $\nu^{1.3}$ power-law
regimes from the stellar wind as plotted in Fig.~\ref{fig:sed} was
noted in Cox et al.\ (1995) and B05 and is required to explain the
observed SED at mm wavelengths.  They proposed that this transition
occurs due to changes in the ionisation properties of the wind but the
location of this transition is not well constrained and consequently
affects the amount of free-free emission expected at submm
wavelengths. For example, if the two power-law regimes turned-over at
870\,$\mu$m instead of 1.2\,mm as plotted here, this would suggest
that 100\,per\,cent of the 870\,$\mu$m flux and 65\,per\,cent of the
450\,$\mu$m could originate from a free-free source.  Changes to the
wavelength of the transition will change the derived properties which
fit the thermal dust SED ($\rm T$, $\beta$) and the dust mass, but
these changes  are well within the errors quoted above.  The
exact location of this transition, though important to understand for
physical reasons, does not affect the dust mass estimated in the
current work.

With gas-to-dust ratio $\sim$100 (Smith \& Ferlund 2007), we estimate
that $>$$\rm 40\,M_{\odot}$ of gas has been ejected. This is
approximately half of the value quoted in G06 but three times more
than the maximum values quoted from previous IR data (Morris et al.\
1999; Smith et al.\ 2003).  Our estimate is also consistent with the
upper end of the mass range ($\rm 15$--$35\rm \,M_{\odot}$) found by Smith
\& Ferlund (2007) using an independent, theoretical model, which
determined the gas mass within the Homunculus using the recent
detection of $\rm H_2$ to determine the density of the gas.  Note that
the dust mass quoted here is different to G06, since the submm flux
attribruted to dust emission has decreased in this work, yet the
dust temperature required to fit the steeper slope between 175\,$\mu$m
and 1000\,$\mu$m is colder here, thereby increasing the mass.

%

\subsection{Mass Loss traced by the Submillimetre Emission}

The \laboca~submm emission is interior to the X-ray shell (tracing
blast wave material), the O\,{\sc III} veil (tracing the recent N-rich
mass loss interacting with older O-rich material (Smith et al.\ 2005)
and the fast blast wave from recent ejecta interacting with the older
N-rich material (Smith 2008).  Since we have little information about
the true structure of the submm source, the \laboca~emission is most
likely a complex combination of the mass lost during the great
ejection phases over the last few hundred years, and the pre-outburst
stellar wind material ejected over the last few thousand years.  From
our constraints on the spatial location of the dust, we therefore
estimate an {\it average} mass loss rate of $>$$10^{-2}\rm
\,M_{\odot}$\,yr$^{-1}$, at least one order of magnitude higher than
suggested in Hillier et al.\ (2001) and Smith et al.\ (2005).  We are
currently unable to resolve the features within the Homunculus and pin
down the exact distribution or contributions of the free-free and dust
components, hence our mass loss has a large associated uncertainty. We
do not know the exact timescale over which the mass has been lost nor
the composition and radiation properties of the dust grains. The
longer timescale put on the submm mass loss in this work compared to
G06, eases the strain on the available mass reservoir of $\eta$ Car
(Smith 2009b), since the $40\,\rm M_{\odot}$ of gas traced by the
submm is likely a combination of eruptions and the stellar
wind. However, we note that such a huge mass loss does provide support
for the idea of dredge-up SN-like events
and also hints that pre-SN mass loss could be significant contributors
to the interstellar dust budget in the early Universe (Morgan \&
Edmunds 2003; Dwek, Galliano \& Jones 2007).

\section{Conclusions} 
\label{sec:conc}

We found that the submm extended emission surrounding $\eta$ Car,
originally identified in G06, is unresolved by the \laboca~beam at
870\,$\mu$m, and is highly variable, following the well-documented
cycle seen at mm and radio wavelengths.  Accounting for the flux
difference from G06 and associating the 800-1000\,$\mu$m with
free-free emission rather than thermal emission from grains, we revise
the dust mass in the stellar wind to $0.4\pm 0.1\,\rm M_{\odot}$. This
translates to a mass loss greater than $\rm 40\,M_{\odot}$ over the
past 1000\, years.  Future observations with Herschel and ALMA are
needed to constrain the submm and mm variability and multi-epoch data
will allow us to separate out the dust and free-free components
throughout the cycle. Higher angular resolution data are needed to
separate out the multiple phases of mass loss, this will provide
crucial information about the evolutionary path to massive-star SN explosions.

\section*{Acknowledgements}
We thank the anonymous referee for insightful comments which greatly
improved this work.  This publication is based on data acquired with
APEX (project 078.F-9036(A) which was a joint science verification
project with 078.F-9017, PI Kramer), a collaboration between
the Max-Planck-Institut fur Radioastronomie, the European Southern
Observatory, and the Onsala Space Observatory. We thank the APEX staff
for their help with observing and data analysis.  We thank Nathan Smith and Kate Brooks
for providing the H$\alpha$ and SIMBA maps.  HLG acknowledges support
from LCOGT.

\end{document}